\documentstyle[prd,aps,epsf]{revtex}
\tighten
\begin{document}
\draft

\preprint{cond-mat/9807164 Imperial/TP/97-98/59}

\twocolumn[\hsize\textwidth\columnwidth\hsize\csname @twocolumnfalse\endcsname

\title{Slow $^{4}He$ Quenches Produce Fuzzy, Transient Vortices}
\author{R.\ J.\ Rivers}
\address{Blackett Laboratory, Imperial College, London SW7 2BZ}
\date{\today}
\maketitle

\begin{abstract}
We examine the Zurek scenario for the production of vortices in
quenches of liquid $^{4}He$  in the light of recent experiments.
Extending our previous results to later times, we argue that short
wavelength thermal fluctuations make vortices poorly defined until
after the transition has occurred. Further, if and when vortices
appear, it is plausible that that they will decay faster than
anticipated from turbulence experiments, irrespective of quench
rates.

\end{abstract}

\pacs{PACS Numbers : 11.27.+d, 05.70.Fh, 11.10.Wx, 67.40.Vs}

\vskip2pc]
Several years ago Kibble initiated a programme\cite{kibble1} for
examining the consequences of phase transitions in the early
universe for large scale structure formation. In particular,
cosmic strings produced at these transitions were seen to have the
potential for creating the observed fluctuations in the microwave
background. Using simple causal arguments\cite{kibble2} he
obtained bounds on the density of strings produced in such
transitions. Unfortunately such consequences can only be inferred
indirectly and with ambiguity. On the other hand, the production
and measurement of vortices in condensed matter (the counterpart
of cosmic strings in QFT) is a viable experiment. In a series of
papers Zurek\cite{zurek1} made predictions for the density of
vortices produced during quenches of liquid $^{4}He$ and $^{3}He$
into their superfluid states, using similar causal
arguments.

Zurek's original assumption was that, as we approach the
(continuous) transition, the correlation length $\xi (t)$ of the
field initially matches the equilibrium correlation length
$\xi_{eq}(t)$ of the field. However, once $\xi (t)$ is growing at
the speed of (second) sound\footnote{The corresponding argument in
QFT replaces the speed of sound by the speed of light.}, at some
time ${\bar t}$ before the transition begins (as characterised by
the divergence of $\xi_{eq}(t)$) it freezes. Only after the
transition has begun does the field unfreeze, at a comparable time
interval ${\bar t}$.

With superfluid $He$ in mind we assume a time-dependent
Landau-Ginzburg free energy of the form
\[
F(t) = \int d^{3}x\,\,\bigg(\frac{-\hbar^{2}}{2m}|\nabla\phi |^{2}
+\alpha (t)|\phi |^{2} + \frac{1}{4}\beta |\phi |^{4}\bigg).
\]
In $F(t)$, $\phi = (\phi_{1} + i\phi_{2})/\sqrt{2}$ is the complex
order-parameter field, whose magnitude determines the superfluid
density.   In equilibrium at temperature $T$, in a mean field
approximation,  the chemical potential $\alpha (T/T_c )$ takes the
form $\alpha (T/T_c ) = \alpha_{0}\epsilon (T/T_c )$, where
$\epsilon = (T/T_{c} -1)$ measures the critical temperature
$T_{c}$ relative to $T$. At a quench the change in $T/T_c$ with
time leads most simply to $\epsilon$ of the form
\[
\epsilon (t) = \epsilon_{0} - \frac{t}{\tau_{Q}}\theta (t)
\]
for $-\infty < t < \tau_{Q}(1 + \epsilon_{0})$, after which
$\epsilon (t) = -1$.  $\epsilon_{0}$ measures the original
relative temperature and $\tau_{Q}$ defines the quench time.  The
quench begins at time $t = 0$ and the transition from the normal
to the superfluid phase begins at time $t =t_{0} =
\epsilon_{0}\tau_{Q}$.

With $\xi_{0}^{2} = \hbar^{2}/2m\alpha_{0}$ and $\tau_{0} = \hbar
/\alpha_{0}$ setting the fundamental distance and time scales,
mean-field critical behaviour fixes\cite{zurek1} ${\bar t} =
\sqrt{\tau_{0}\tau_{Q}}$.  The frozen correlation length of the
field ${\bar\xi}= \xi_{0}(\tau_{Q}/\tau_{0})^{1/4}\approx
\xi_{eq} (t_{0}-{\bar t})\approx\xi_{eq}
(t_{0}+{\bar t})$.
The ordering of the field on large scales will be marred by
topological defects, in this case vortices.
 If the {\it initial} vortex
separation is $\xi_{def}$, then their density, $n_{def}$, is
$n_{def} = O(1/\xi_{def}^{2})$. Zurek initially made the {\it second}
assumption (matched by Kibble in QFT) that $\xi_{def}=
O({\bar\xi})$ whereby
\[
n_{def} =
\frac{1}{f^{2}\xi_{0}^{2}}\sqrt{\frac{\tau_{0}}{\tau_{Q}}},
\]
where $f = O(1)$. Since $\xi_{0}$ also measures cold vortex
thickness, $\tau_{Q}\gg \tau_{0}$ corresponds to a measurably
large number of widely separated vortices.

This is a remarkable prediction in that it says that, in the first instance, the
dimensionful interactions $\beta |\phi |^{4}$ introduce no new scale.
It has been the motive for several experiments.
One\cite{lancaster} at Lancaster on vortex production in
$^{4}He$ shows consistency with the prediction, whereas a second
experiment\cite{lancaster2} on $^{4}He$ finds no vortices.  On the
other hand, vortex production in quenches of $^{3}He$ into its
$B$-phase in two separate experiments at Grenoble\cite{grenoble}
and Helsinki\cite{helsinki} show quantitative agreement with
$n_{def}$ above. This is even though it is now argued\cite{kopnin}
that the Helsinki experiment should {\it not} show agreement
because of the geometry of the heating event.

In an earlier letter\cite{ray2} we argued that, RG improvement
apart, the simple prediction above for $n_{def}$ is unreliable on
two counts. Firstly, and crucially, vortex separation cannot be
related simply to field or phase correlation length.  It is much
more sensible to relate vortex density to the density of line
zeroes, $n_{zero}$, of the field, since a classical vortex is a
tube of false vacuum, with a line zero at its core. For Gaussian
field fluctuations
\[
\langle\phi_{a} ({\bf r},t)\phi_{b} ({\bf 0},t)\rangle
=\delta_{ab}G({\bf r},t),
\]
 the line-zero density $n_{zero}(t)$ can be shown
\cite{halperin,maz} to have the form (where primes denote
differentiation with respect to $r$)
\[
n_{zero}(t) = \frac{-1}{2\pi}\frac{G\prime\prime (0,t)}{G(0,t)}.
\]
It depends manifestly  on the {\it short-distance} behaviour of
$G(r,t)$, rather than the correlation length ${\bar\xi}$ that
follows from large-distance behaviour.  Secondly, if a line zero
is to be counted as  a vortex, at the very least it should have
fractal dimension close to unity on a scale comparable to the
classical vortex thickness $\xi_{0}$, even though it may behave
like a random walk on larger scales. Otherwise the density would
depend on the scale at which we look. This is equally true for
cosmic strings\cite{ray}.

In \cite{ray2} we analysed the model until the time that the
transition was effected. In this letter we shall now argue that,
even when the transition is complete, for slow quenches the short
wavelength thermal fluctuations are likely to persist for some
time at a level that prevents an approximately scale-invariant
definition of vortex density. Further, after that time it is
plausible that such vortices as do exist decay more rapidly than
is assumed in the analysis of \cite{lancaster2}.

Motivated by Zurek's later numerical\cite{laguna,zurek2}
simulations, we adopt the time-dependent Landau-Ginzburg (TDLG)
equation for $F$,
\[
\frac{1}{\Gamma}\frac{\partial\phi_{a}}{\partial t} =
-\frac{\delta F}{\delta\phi_{a}} + \eta_{a},
\]
where $\eta_{a}$ is Gaussian thermal noise, satisfying
\[
\langle\eta_{a} ({\bf x},t)\eta_{b} ({\bf y}',t')\rangle= 2\Gamma
T(t)\delta_{ab}\delta ({\bf x}-{\bf y})\delta (t -t').
\]

In space, time and temperature units in which $\xi_{0} = \tau_{0}
= k_{B} =1$, the TDLG equation becomes
\[
{\dot\phi}_{a}({\bf x},t) = - [-\nabla^{2} + \epsilon
(t)+{\bar\beta}|\phi ({\bf x},t)|^{2}]\phi_{a} ({\bf x},t)
+{\bar\eta}_{a} ({\bf x},t),
\]
where ${\bar\beta}$ is the rescaled coupling and ${\bar\eta}$ the
rescaled noise. We  adopt the approximation of preserving Gaussian
fluctuations by linearising the self-interaction as
\[
{\dot\phi}_{a}({\bf x},t) = - [-\nabla^{2} + \epsilon_{eff}
(t)]\phi_{a} ({\bf x},t) +{\bar\eta}_{a} ({\bf x},t),
\]
where $\epsilon_{eff} $ contains a (self-consistent) term
$O({\bar\beta}\langle |\phi |^{2}\rangle)$.
Additive renormalisation is necessary, so
that $\epsilon_{eff}\approx \epsilon$, as given earlier, for
$t\leq t_{0}$.

Self-consistent linearisation is the standard approximation in
non-equilibrium QFT\cite{boyanovsky}, but is not strictly
necessary here, since numerical simulations that identify line
zeroes of the field can be made that use the full
self-interaction\cite{zurek2}. However, to date none address
the questions we are posing here exactly,
and until then there is virtue in analytic approximations provided
they are not taken too seriously.

 The solution for $G(r,t)$ from the
linear equation for $\phi_{a}$ is straightforward, most usefully
in the Schwinger proper-time representation, as $G(r,t) =$
\[
\int_{0}^{\infty} d\tau \,{\bar T}(t-\tau /2)
\bigg(\frac{1}{4\pi\tau}\bigg)^{3/2}
e^{-r^{2}/4\tau}\,e^{-\int_{0}^{\tau} ds\,\,\epsilon_{eff} (t-
s/2)},
\]
where  ${\bar T}$ is the rescaled temperature.

 For constant $\epsilon_{eff}$, as happens at early times, we
recover the usual Yukawa correlator. However, a saddle-point
calculation at time $t_{0}$ gives $\xi (t_{0}) = O({\bar\xi})$,
 confirming Zurek's causal result for the freezing of the field correlation
 length.

A direct substitution of $G(r,t)$ into $n_{zero}$ is impossible
because of UV singularities at small $\tau$. However, a
starting-point for counting vortices in superfluids is to count
line zeroes of an appropriately coarse-grained field, in which
structure on a scale smaller than $\xi_{0}$, the classical vortex
size, is not present.
This is, indeed, the basis of Zurek's and other
numerical simulations. For the moment, we put in a momentum cutoff
$k^{-1}> l =\bar{l}\xi_{0} =O(\xi_{0} )$ by hand,
which corresponds to damping the singularity  in $G(r,t)$ at
$\tau = 0$ as
\[
G_{l}(r,t)=\int_{0}^{\infty}\frac{ d\tau\,{\bar T}(t-\tau
/2)}{[4\pi (\tau + {\bar l}^{2})]^{3/2} }
e^{-r^{2}/4\tau}\,e^{-\int_{0}^{\tau} ds\,\,\epsilon_{eff} (t-
s/2)},
\]
making $G_{l}(0,t)$ finite. We stress that, for $t\approx t_{0}$,
the correlation length ${\xi}$ remains $O({\bar\xi})$, {\it
independent} of $l$.

Assuming a {\it single} zero of $\epsilon_{eff} (t)$ at $t =
t_{0}$, at $r=0$ the exponential in the integrand peaks at $\tau
={\bar\tau} = 2(t-t_{0})$.  Expanding about ${\bar\tau}$ to
quadratic order gives
\[
G_{l}(0,t)\approx {\bar
T}_{c}\,e^{2\int_{t_{0}}^{t}du\,|\epsilon_{eff} (u)|
}\int_{0}^{\infty} \frac{ d\tau\,e^{-(\tau -
2(t-t_{0}))^{2}|\epsilon '(t_{0})|/4}}{[4\pi (\tau + {\bar
l}^{2})]^{3/2} } .
\]
The effect of the back-reaction is to stop the growth of
$G_{l}(0,t)-G_{l}(0,t_{0})= \langle |\phi |^{2}\rangle_{t}-\langle
|\phi |^{2}\rangle_{0}$ at its symmetry-broken value
${\bar\beta}^{-1}$ in our dimensionless units.
 We see that this
cessation of growth requires
$\lim_{u\rightarrow\infty}\epsilon_{eff} (u) = 0$. That is, we
must choose
\[
\epsilon_{eff}(t) = \epsilon (t)  +
{\bar\beta}(G_{l}(0,t)-G_{l}(t_{0},0)),
\]
thereby preserving Goldstone's theorem.

Beyond that, what is remarkable in this approximation is that the
density of line zeroes uses {\it no} property of the self-mass
contribution to $\epsilon_{eff}(t)$, self-consistent or otherwise.
In $n_{zero}$ all prefactors cancel\footnote{Our ignoring prefactors
in \cite{ray2} was fortuitous, leaving our conclusions obtained
there unaffected.}, to give
\[
n_{zero}(t) = \frac{1}{4\pi}\frac{\int_{0}^{\infty}
\frac{d\tau}{(\tau + {\bar l}^{2})^{5/2}} \,e^{-(\tau -
2(t-t_{0}))^{2}/4{\bar t}^{2}}} {\int_{0}^{\infty}
\frac{d\tau}{(\tau + {\bar l}^{2})^{3/2}} \,e^{-(\tau - 2(t-
t_{0}))^{2}/4{\bar t}^{2}}}
\]
on using the definition $\tau_{Q} = {\bar t}^{2}$ in natural
units. At $t=t_{0}$ both numerator and denominator are dominated
by the short wavelength fluctuations at small $\tau$. Even though
the field is correlated over a distance ${\bar\xi}\gg l$ the
density of line zeroes $n_{zero} = O(l^{-2})$ depends entirely on
the scale at which we look. However, as time passes the peak of
the exponential grows and $n_{zero}$ becomes increasingly
insensitive to $l$. How much time we have depends on the magnitude
of ${\bar\beta}$, since once $G(0,t)$ has reached this value it
stops growing. Since $G(0,t) = O(\exp(((t-t_{0})/{\bar t})^{2})$
at early times the backreaction is implemented extremely rapidly.
We can estimate the time $t^{*}$ at which this happens by
substituting $\epsilon (u)$ for $\epsilon_{eff} (u)$ in the
expression for $G_{l}(0,t)$ above.

For $t>t^{*}$ the equation for $n_{zero}(t)$ is not so simple
since the estimate above, based on a single isolated zero of
$\epsilon_{eff} (t)$, breaks down because of the approximate
vanishing of $\epsilon_{eff} (t)$ for $t>t^{*}$.  A more careful
analysis shows that $G_{l}(0,t)$ can be written as
\[
G_{l}(0,t)\approx \int_{0}^{\infty} \frac{ d\tau\,{\bar T}(t-\tau
/2)}{[4\pi (\tau + {\bar l}^{2})]^{3/2} } {\bar G}(\tau,t),
\]
where ${\bar G}(\tau,t)$ has the same peak as before at $\tau =
2(t-t_{0})$, in the vicinity of which
\[
{\bar G}(\tau,t) = e^{2\int_{t_{0}}^{t}du\,|\epsilon_{eff}
(u)|}\,e^{-(\tau - 2(t-t_{0}))^{2}/4{\bar t}^{2}},
\]
but ${\bar G}(\tau,t)\cong 1$ for $\tau < 2(t-t^{*})$. Thus, for
$\tau_{Q}\gg\tau_{0}$, $G_{l}(0,t)$ can be approximately separated
as
\[
G_{l}(0,t)\cong G_{l}^{UV}(t) + G^{IR}(t),
\]
where
\[G_{l}^{UV}(t)= {\bar T}(t)\,\int_{0}^{\infty}
d\tau\,/[4\pi (\tau + {\bar l}^{2})]^{3/2}
\]
 describes  the
scale-{\it dependent} short wavelength thermal noise, proportional
to temperature, and
\[
G^{IR}(t) =\frac{{\bar T}_{c}}{(8\pi(t-t_{0}))^{3/2}}
\,\int_{-\infty}^{\infty}d\tau {\bar G}(\tau,t)
\]
describes the scale-{\it independent}, temperature independent,
long wavelength fluctuations. A similar decomposition
$G\prime\prime_{l}(0,t)\cong G\prime\prime_{l}^{UV}(t) +
G\prime\prime^{IR}(t)$ can be performed in an obvious way, with
$G\prime\prime^{IR}(t)/G^{IR}(t)= O(t^{-1})$.

Firstly, suppose that, for $t\geq t^*$,
  $ G^{IR}(t)\gg G_{l}^{UV}(t)$ and  $
G\prime\prime^{IR}(t)\gg G\prime\prime_{l}^{UV}(t)$, as would be
the case for a temperature quench ${\bar T}(t)\rightarrow 0$.
 Then, with little thermal noise, we have widely separated
line zeroes, with density $n_{def}(t)\approx
-G\prime\prime^{IR}(t)/2\pi G^{IR}(t)$. With $\partial
n_{zero}/\partial l$ small in comparison to $n_{zero}/l$ at $l =
\xi_{0}$ we identify such essentially non-fractal line-zeroes with
vortices, and $n_{zero}$ with $n_{def}$.  Of course, we require
non-Gaussianity to create true classical energy profiles.
Nonetheless, the Halperin-Mazenko result may be well approximated
even when the fluctuations are no longer Gaussian\cite{calzetta}.
This is supported by the observation that, once the line zeroes
have straightened on small scales at $t>t^*$, the Gaussian field
energy, largely in field gradients, is
\[
{\bar F}\approx\langle\int_{V}
d^{3}x\,\frac{1}{2}(\nabla\phi_{a})^{2}\rangle= -VG''(0,t),
\]
where $V$ is the spatial volume. This matches the energy
\[
{\bar E}\approx V n_{def}(t)(2\pi G(0,t)) = -VG''(0,t)
\]
possessed by a network of classical global strings with density
$n_{zero}$, in the same approximation of cutting off their
logarithmic tails.

 For times $t>t^{*}$, but not too late,
\[
n_{def}(t)\approx \frac{{\bar t}}{8\pi (t-t_{0})
}\frac{1}{\xi_{0}^{2}}\sqrt{\frac{\tau_{0}}{\tau_{Q}}},
\]
the solution to Vinen's equation\cite{vinen}
\[
\frac{\partial n_{def}}{\partial t} = -\chi_{2}\frac{\hbar}{m}
n_{def}^{2},
\]
where $\chi_{2} = 4\pi$ is a {\it universal constant}.

 This decay law is assumed in the analysis of the Lancaster
experiments, in which the density of vortices is inferred from the
intensity of the signal of scattered second sound. RG improvement
leaves $^{3}He$ unchanged, but for $^{4}He$ it redefines
$\chi_{2}$ to $\chi_{2}(1-T/T_{c})^{-1/3}
> \chi_{2}$. The problem is that this makes an already large $\chi_2$
even larger. In the attenuation of second sound the signal to
noise ratio is approximately $O(1/\chi_{2}t)$. The empirical value
of $\chi_2$ used in the Lancaster experiments is not taken from
quenches, but turbulent flow experiments. It is
suggested\cite{lancaster2} that $\chi_{2} \approx 0.005$, a good
three orders of magnitude smaller than our prediction above.
Although the TDLG theory is not very reliable for $^{4}He$, if our
estimate is sensible it does imply that vortices produced in a
{\it temperature} quench decay much faster than those produced in
turbulence.
$^{3}He$
experiments provide no check.

In reality, the situation for the Lancaster $^{4}He$ experiments
is even more complex, since they are {\it pressure} quenches for
which the temperature $T$ is almost {\it constant} at $T\approx
T_{c}$.
 Unlike
temperature quenches\cite{zurek2,boyanovsky}, thermal fluctuations
here remain at full strength. Even for $^3 He$, $T/T_{c}$ never
gets very small, and henceforth we take $T=T_{c}$ in $G_{l}(0,t)$
above. The necessary time-{\it independence} of $ G^{IR}(t)$ for
$t>t^*$ is achieved by taking $\epsilon_{eff} (u)= O(u^{-1})$. In
consequence, as $t$ increases beyond $t^{*}$ the relative
magnitude of the UV and IR contributions to $G_{l}(0,t)$  remains
{\it approximately constant}. Further, since for $t = t^*$,
\[
e^{2\int_{t_{0}}^{t}du\,|\epsilon_{eff} (u)|}\,e^{-(\Delta
t)^{2}/{\bar t}^{2}}\approx 1,
\]
this ratio is the ratio at $t = t^*$.

Nonetheless, as long as the UV fluctuations are insignificant at
$t=t^*$ the density of line zeroes will remain largely independent
of scale. This follows if $ G\prime\prime^{IR}(t^*)\gg
G\prime\prime_{l}^{UV}(t^*)$, since $G\prime\prime_{l}(0,t)$
becomes scale-independent later than $G_{l}(0,t)$. In \cite{ray2}
we showed that this is true provided
\[
(\tau_{Q}/\tau_{0})(1-T_{G}/T_{c})<C\pi^{4},
\]
where $C= O(1)$ and $T_G$ is the Ginzburg temperature. With
$\tau_{Q}/\tau_{0} = O(10^{3})$ and $(1-T_G /T_c ) = O(10^{-12})$
this inequality is well satisfied for a linearised TDLG theory for
$^{3}He$ derived\footnote{Ignoring the position-dependent
temperature of \cite{kopnin}} from the full TDGL
theory\cite{Bunkov}, but there is no way that it can be satisfied
for $^{4}He$, when subjected to a slow mechanical quench, as in
the Lancaster experiment, for which $\tau_{Q}/\tau_{0} =
O(10^{10})$, since the Ginzburg regime is so large that $(1-T_G
/T_c ) = O(1)$. Effectively, the $^{4}He$ quench is {\it nineteen}
orders of magnitude slower than its $^{3}He$ counterpart. It is
satisfying to find the Ginzburg regime,
originally\cite{kibble1,copeland} but erroneously thought to
determine the initial defect density, but then
discarded\cite{kibble2,zurek1}, reappear as an essential
ingredient in undermining universality.

Thus, for $^{3}He$, with negligable UV contributions, we estimate
the primordial density of vortices as
\[
n_{def}(t^{*})\approx \frac{{\bar t}}{8\pi (t^{*} - t_{0})
}\frac{1}{\xi_{0}^{2}}\sqrt{\frac{\tau_{0}}{\tau_{Q}}},
\]
in accord with the original prediction of Zurek. Because of the
rapid growth of $G(0,t)$, $(t^{*}-t_{0})/{\bar t} =p > 1 = O(1)$.
With $p$ behaving as $(\ln{\bar\beta})^{1/2}$ there is very little
variation. For $^{3}He$ quenches $p\approx 5$ (and for $^{3}He$
quenches $p\approx 3$).
 We note that the factor\footnote{An errant factor of 3 appeared
 in the result of
\cite{ray2}} of $f^{2}=8\pi p$ gives a value of $f = O(10)$, in
agreement with the empirical results of \cite{grenoble} and the
numerical results of \cite{laguna,zurek2}\footnote{The temperature
quench of the latter is somewhat different from that considered
here, but should still give the same results in this case}.

However, if the inequality is badly violated, as happens with
$^{4}He$ for slow pressure (but not temperature) quenches, then
the density of zeroes $n_{def}= O(l^{-2})$ after $t^*$ depends
exactly on the scale $l$ at which we look and they are not
candidates for vortices. Since the whole of the quench takes place
within the Ginzburg regime this is not implausible. However, it is
possible that, even though the thermal noise never switches off,
there is no more than a postponement of vortex production, since
our approximations must break down at some stage. The best outcome
is to assume that the effect of the thermal fluctuations on
fractal behaviour is diminished, only leading to a delay in the
time at which vortices finally appear. Even if we suppose that
$n_{def}$ above is a starting point for calculating the density at
later times, albeit with a different $t_{0}$, thereby preserving
Vinen's law, we then have the earlier problem of the large
$\chi_{2} = O(f^2 )$. In the absence of any mechanism to reduce
its value drastically, this would make it impossible to see
vortices. As a separate observation, we note that the large value
of $f^2$ in the prefactor of $n_{def}$ is, in itself, almost
enough to make it impossible to see vortices in $^{4}He$
experiments, should they be present. This will be pursued
elsewhere.

In summary, this work suggests that for slow pressure quenches in
$^{4}He$, we see no well-defined vortices at early times because
of thermal fluctuations, and it is plausible that, if we do see
them at later times, there are less than we would have expected
because of their rapid decay and their initial low density. The
situation is different for $^3 He$. That the density of such
vortices as appear may agree with the Zurek prediction is
essentially a consequence of dimensional analysis, given that the
main effects of the self-interaction have a tendency to cancel in
the counting of vortices, without introducing new scales.

 This work is
the result of a network supported by the European Science
Foundation.  I thank Tom Kibble, Grisha Volovik, Nuno Antunes and
Eleftheria Kavoussanaki for helpful conversations and analysis.

\end{document}